# Causal Inference with MNAR Self-Masking Confounders: A Stratified Delta-Imputed Propensity Estimation Method

Md. Niamul Islam Sium[1]; Mohammad Hridoy Patwary[2*]

[1] Mathematical Sciences, The University of Texas at El Paso, 500 W. University Ave, El Paso, Texas, USA.

[2] Health Systems and Population Studies Division, icddr,b, GPO Box 128, Dhaka-1000, Dhaka, Bangladesh.

*Corresponding author:

Mohammad Hridoy Patwary

Email: mpatwary@isrt.ac.bd

## 1. Summary

In observational studies, causal inference becomes difficult when confounders are missing-not-at-random (MNAR), particularly where the missingness depends on the confounder's own unreported value (self-masking). Existing methods for handling MNAR confounders often rely on strong, unverifiable assumptions, leading to biased estimates. We propose a simple approach with Stratified Delta-Imputed Propensity Estimator (SDIPE) in the presence of self-masking confounders. SDIPE first stratifies data into observed and missing groups, imputes missing confounders via delta-adjusted multiple imputation. Then, within each group, average-treatment-effects (ATEs) are estimated by stabilized-inverse-probability-weights. The final ATE is obtained by combining the subgroup-specific estimates, weighted by respective proportions in the sample.

Simulation study shows that SDIPE achieves low bias and near-nominal coverage (~94–96%) across varying missingness, sample sizes, and treatment prevalence. In contrast, conventional sensitivity-based multiple imputation exhibits substantial bias and poor coverage (18-89%). Additionally, SDIPE is robust to the choice of the delta parameter. Applied to NHANES-2017–2018, SDIPE estimates that married individuals have a 1.19-point lower depression score than unmarried individuals (95%CI: -1.76, -0.64), adjusting for MNAR income data. SDIPE provides a practical and robust approach for causal inference with self-masking MNAR confounders, offering improved performance over existing methods without requiring restrictive assumptions about the missingness mechanism.

## 2 Introduction

Observational studies have become indispensable for evaluating causal effects when randomized experiments are ethically or practically infeasible (Gianicolo et al., 2020). Estimating causal effects using observational data is, however, challenging due to the need for adequate control of confounding variables that jointly influence both treatment assignment and the outcome. Established methods like parametric g-formula (Lodi et al., 2015; Taubman et al., 2009), propensity score-based methods (Cain et al., 2010; Hernán et al., 2000; Neugebauer et al., 2014), or doubly robust methods (Bang & Robins, 2005; Tran et al., 2019; Wen et al., 2023) are widely used when confounders are all completely observed, commonly referred to as the exchangeability assumption (M. A. Hernán & J. M. Robins, 2020).

In reality, this assumption is often violated. Key confounders may be missing due to various reasons, and this missingness of confounders often plays a crucial role in causal effect estimation; if unaddressed, it can result in substantial bias in treatment effect estimates (Yang et al., 2019). Different approaches are there to handle unmeasured confounders: propensity score calibration (Stürmer et al., 2005), sensitivity analysis (VanderWeele & Arah, 2011), E-value approach (VanderWeele & Ding, 2017). But partially missing confounders is often an overlook detail in causal inference, which seems very common in observational data (Lu & Ashmead, 2018).

These missing values in confounders may arise under several missing mechanisms: missing completely at random (MCAR), missing at random (MAR), and missing not at random (MNAR) (Little & Rubin, 2019). If a confounder is MCAR, then complete case analysis can still yield an unbiased estimate of the average causal effect (Imai & van Dyk, 2004). When confounders are missing at random (MAR), commonly used methods include multiple imputation (MI) (Mitra & Reiter, 2011; Seaman & White, 2014) and inverse probability weighting (IPW) (Ross et al., 2021). Methods for handling MNAR confounders have been developed recently. For instance, Ding and Geng (Ding & Geng, 2014) examined the identifiability of causal effects in randomized experiments including MNAR covariates only for discrete covariates and outcomes, later Corder and Yang (Corder & Yang, 2020) demonstrated that causal effects can be identified under a specific assumption: the missingness mechanism is independent of the outcome. To estimate causal effects under this framework, they proposed both a nonparametric two-stage least squares estimator and parametric likelihood-based methods. Additionally, Yang et al. (Yang et al., 2014) addressed nonignorable missing covariates in the context of instrumental variable approaches for unmasured confounding. They developed a maximum likelihood estimation method using the EM algorithm to produce unbiased causal effect estimates. While these methods offer valuable insights, they often rely on additional assumptions that cannot be verified from the observed data. Moreover, none of the existing approaches has explicitly addressed scenarios involving the self-masking (SM) MNAR mechanism.

SM is one kind of MNAR missing mechanism where the probability of being missing depends on the unobserved value itself (Little & Rubin, 2019), as seen in economics (high-income individuals omitting financial data; (Slemrod, 2007)), education (low-performing students' test scores withheld; (Van der Linden & van der Linden, 2016)). Methodologically, SM creates a perfect storm of identifiability challenges. When missingness depends solely on the unobserved confounder (without auxiliary predictors), the full data distribution becomes non-identifiable without strong assumptions (Mohan & Pearl, 2021), which poses challenges to estimate the causal effect. To bridge this gap, we introduce the Stratified Delta-Imputed Propensity Estimator (SDIPE) that integrates delta-adjusted MI with stratified propensity score estimation. Unlike existing methods that require strong assumptions about the independence of the missingness mechanism from the treatment or outcome, SDIPE adopts a flexible imputation-based framework with delta-adjustment for sensitivity analysis, thereby offering broader practical applicability.

## 3 Material and Methods

### 3.1 Causal Framework & Notations

Consider, we have $n$ independent samples. Assume a continuous outcome variable $Y$, a binary exposure of interest, $A$ and a vector of continuous potential confounders, $\boldsymbol{C} = (C_1, C_2, \ldots, C_p)^T$.

Under the potential outcome framework, each individual in the population of interest has a potential outcome ($Y^a$) under each treatment state ($A = a; a \in \{0,1\}$), even though each individual can be observed in only one treatment state at any point in time (Pearl, 2010). The consistency assumption links this observed outcome to the relevant potential outcome (Hernán & Robins, 2010): if an individual receives treatment level $A_i = a$, then the observed outcome $Y_i$ equals the potential outcome under that treatment level, i.e.,

$$\text{if } A_i = a \text{ then, } Y_i^a = Y_i^A = Y_i.$$

Using this notion, the observed outcome $Y$ can be written as

$$Y = A \times Y^{a=1} + (1 - A) \times Y^{a=0},$$

where,

$Y^{a=1}$ is the potential outcome under exposed condition,

$Y^{a=0}$ is the potential outcome under unexposed condition.

### 3.2 Missing Data Structure

Let, $C = (Z, C_2, C_3, \ldots, C_p)' = (Z, W)'$ denote the vector of confounders, where $W$ is fully observed, but $Z$ is partially observed. Let $R_Z \in \{0,1\}$ be the missingness indicator for $Z$, with $R_z = 1$ if $Z$ is observed and $R_z = 0$ if $Z$ is missing.

We assume a SM MNAR mechanism, meaning the probability that $Z$ is observed depends only on its own value, i.e. $P(R_Z = 1|Z, W, A, Y) = P(R_z = 1|Z)$.

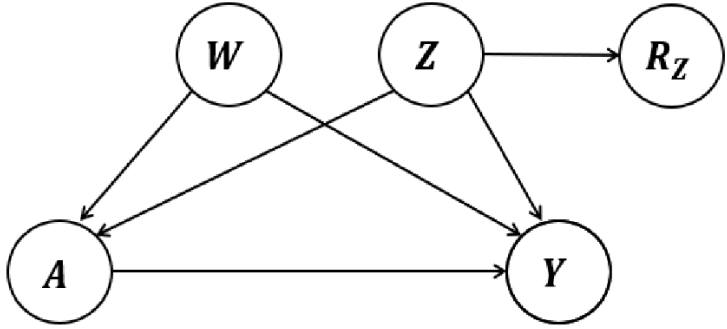

*Figure 1: A DAG presenting the data generating process.*

### 3.3 Additional Assumptions and Decomposition of the ATE

Now, to estimate the average treatment effect (ATE), $\tau = E(Y^{a=1} - Y^{a=0})$, the two key assumptions must hold (M. Hernán & J. Robins, 2020):

1. Positivity: $0 < P(A = a|C = c) < 1 \ \forall \ a \in \{0,1\}$, and $\forall \ c$ with $P(C = c) > 0$
2. Conditional Exchangeability: $Y^a \perp A \mid C$ \qquad (1)

Equation (1) holds if the confounder vector $C$ is fully observed. However, when element(s) of $C$ are partially observed, the equation (1) does not hold as the treatment assignment $A$ may still be associated with the potential outcomes through the unmeasured or missing components of $C$ (M. Hernán & J. Robins, 2020). In this scenario, the ATE can be divided into two parts based on the values of $R_z$ as:

$$\tau = E(Y^{a=1} - Y^{a=0}) = E_{R_z}[E(Y^{a=1} - Y^{a=0}|R_z)]$$

$$= \sum_{r=0}^{1} E(Y^{a=1} - Y^{a=0}|R_z = r) \times P(R_z = r)$$

$$= E(Y^{a=1} - Y^{a=0}|R_z = 1) \times P(R_z = 1) + E(Y^{a=1} - Y^{a=0}|R_z = 0) \times P(R_z = 0), \quad (2)$$

where, the treatment effect from the first part is identifiable, but the second one is not. As a result, estimates of the ATE based on only observed data may be biased, and causal inference is no longer valid without further assumptions or methods.

### 3.4 Stratified Delta-Imputed Propensity Estimator

To address the identification challenges arising from the missingness in the confounder $Z$ under MNAR mechanism, we employed a MI technique combined with sensitivity analysis and subgroup estimation of the ATE. MI combined with sensitivity analysis has been widely recommended for handling missing variables under MNAR assumptions (Carpenter et al., 2023; Rezvan et al., 2018). This approach enables researchers to explore how inferences may vary under plausible departures from the MAR assumption. Sensitivity analysis methods often rely on shifting the distribution of the missing data relative to the observed, using techniques such as location and/or scale adjustments (Dioni et al., 2025). Among these, a location shift ($\delta$) adjustment seems appropriate for our context, given our use of a linear model for imputing the partially missing confounder $Z$. This method assumes that the distribution of the missing values is a shifted version of the observed data, enabling a transparent and interpretable sensitivity analysis under the MNAR framework. It is particularly suitable in linear settings where departures from MAR can be modeled as additive shifts in the mean structure (Gachau et al., 2020).

### 3.5 Estimation Procedure

For simplicity, we consider two continuous scalar confounding variables, $Z$ and $W$, where $Z$ is partially observed and $W$ is fully observed. Let $R_z$ denote the missingness indicator for $Z$, such that $R_z = 1$ if $Z$ is observed and $R_z = 0$ if $Z$ is missing. We assume that the missingness mechanism is SM, meaning that $R_z$ depends only on the value of $Z$ itself and not on any other variables. Under this setting, we observe the vector $O = (A, Y, W, R_z)'$ for all individuals, while the value of $Z$ is available only for those with $R_Z = 1$. Let, $Z^{obs}$ denotes the observed portion of $Z$ (for which $R_Z = 1$) and $Z^{miss}$ denotes the missing portion of $Z$ (for which $R_Z = 0$).

We propose separate estimation of ATE through stabilized IP weights (SW) for missing and observed portion of the data:

$$\hat{\tau} = E(Y^{a=1} - \widehat{Y^{a=0}}|R_z = 1) \times P(R_z = 1) + E(Y^{a=1} - \widehat{Y^{a=0}}|R_z = 0) \times P(R_z = 0)$$

$$= \hat{\tau}^{SW}{}_{R_{z=1}} \times P(R_z = 1) + \hat{\tau}^{SW}{}_{R_{z=0}} \times P(R_z = 0).$$

Here,

$\hat{\tau}^{SW}{}_{R_{z=1}}$: estimated ATE among individuals with observed $Z$, using SW.

$\hat{\tau}^{SW}{}_{R_{z=0}}$: estimated ATE among individuals with missing $Z$, using SW.

Now to obtain $\hat{\tau}^{SW}{}_{R_{z=0}}$, we apply MI to predict the values of $Z^{miss}$ by modeling $Z^{obs}$ as a function of the observed data $(A, W, Y)'$ and then using the fitted model to predict for $Z^{miss}$. A shift parameter $(\delta)$ was used during the imputation to allow for departures from the MAR assumption.

## 4 Simulation Study

We conducted a simulation study to rigorously evaluate the performance of our proposed estimating method for average treatment effects when facing SM missingness in a confounder. The simulation framework was designed to mirror real-world conditions where missingness in confounder $Z$ depends on its own unobserved value, a real challenge in observational studies.

Data generation followed a well-defined causal structure: For each subject $i = 1, \dots, n$.

- We first simulated covariate $W_i \sim N(0,1)$ and the confounder $Z_i \sim N(0,1)$.
- A binary treatment $A_i$ was then generated from a Bernoulli distribution with probability $logit^{-1}(\alpha_1 + \alpha_2 W_i + \alpha_3 Z_i)$, with $\alpha_2 = 0.5, \alpha_3 = 0.5$, and $\alpha_1$, where $\alpha_1$ controlled the overall treatment prevalence in the simulated dataset.
- The continuous outcome $Y_i$ was constructed as $Y_i = \beta_0 + \beta_1 A_i + \beta_2 W_i + \beta_3 Z_i + \epsilon_i$, here $\beta_1 = 1.5$ reflected the true ATE and the values of the other parameters were $\beta_2 = 0.8$, and $\beta_3 = 0.7$.
- The SM mechanism was implemented by setting $Z$ as missing whenever $logit^{-1}(a + Z_i) = 1$, where parameter $a$ controlled missingness probabilities at different percentages.

Our SDIPE method strategically splits data into Observed ($Z^{obs}$) and Missing ($Z^{miss}$) groups. For the missing group, we imputed $Z$ via regressing on $(A, W, Y)$ with incorporating a sensitivity adjustment ($\delta$) to quantify MNAR bias. ATE estimates were derived using SW with the imputed confounder, and final estimates combined group-specific results weighted by sample proportions.

Performance was assessed through three key metrics: average bias, 95% confidence intervals (CIs), and corresponding coverage probabilities. To compute the bias, and CIs, we used 500 Monte Carlo (MC) replicates with 10 imputations each. For assessing coverage probability, we used the same number of MC replicates and imputations, with each replicate including 200 bootstrap samples.

**4.1 Results**

Our proposed estimation method consistently demonstrated superior performance compared to sensitivity-based MI across all tested conditions of missing data and sample sizes. We examined two treatment prevalence settings (20% and 40%), two sample sizes (n = 500 and n = 1000), and three levels of missingness in the partially observed confounder (10%, 30%, and 50%).

Table 1 represents the findings for 20% of overall treatment prevalence, where *n* indicates different sample sizes. For 500 sample size, our proposed method exhibited near-negligible relative bias, ranging from 1.20% to 0.49% across missing data proportions. In contrast, sensitivity-based MI showed several times higher relative bias (5.00% to16.09%). This scenario was the same for 1000 sample size. Sensitivity based MI showed a narrower CI than our proposed method in almost all aspects.

*Table 1 Estimated relative biases, confidence intervals and coverage rates of ATE over different missing percentages of confounder across several sample sizes for 20% of total treatment prevalence.*

| Methods | n | Missing percentage | Bias$^\dagger$ | 95% CI | Coverage probability |
|---|---|---|---|---|---|
| Multiple Imputation | 500 | 10 | 5.48 | [1.3, 1.86] | 0.86 |
|  |  | 30 | 11.08 | [1.37, 1.97] | 0.75 |

| Methods | n | Missing percentage | Bias† | 95% CI | Coverage probability |
|---|---|---|---|---|---|
| with Sensitivity | | 50 | 16.09 | [1.46, 2.05] | 0.59 |
| | | 10 | 5.00 | [1.35, 1.78] | 0.86 |
| | 1000 | 30 | 11.58 | [1.47, 1.88] | 0.54 |
| | | 50 | 15.75 | [1.53, 1.95] | 0.37 |
| SDIPE | | 10 | 1.20 | [1.18, 1.80] | 0.92 |
| | 500 | 30 | 0.33 | [1.12, 1.85] | 0.95 |
| | | 50 | 0.49 | [1.03, 1.99] | 0.93 |
| | | 10 | 0.27 | [1.27, 1.73] | 0.94 |
| | 1000 | 30 | 0.00 | [1.21, 1.76] | 0.94 |
| | | 50 | 0.20 | [1.14, 1.82] | 0.93 |

†Percentage of bias relative to true ATE.

The coverage rate for our proposed method was consistently higher, closely approximating the nominal 95% target regardless of missingness or sample size. Conversely, sensitivity-based MI showed marked declines in coverage under higher missingness: coverage probability fell to 0.59 (n = 500) and 0.37 (n = 1000) at 50% missing data.

*Table 2 Estimated relative biases, confidence intervals and coverage rates of ATE over different missing percentages of confounder across several sample sizes for 40% of total treatment prevalence.*

| Methods | n | Missing percentage | Bias† | 95% CI | Coverage probability |
|---|---|---|---|---|---|
| | 500 | 10 | 4.65 | [1.39, 1.77] | 0.89 |

| | | 30 | 9.58 | [1.43, 1.84] | 0.71 |
| --- | --- | --- | --- | --- | --- |
| Multiple Imputation with Sensitivity | | 50 | 14.78 | [1.50, 1.96] | 0.46 |
| | 1000 | 10 | 5.05 | [1.43, 1.72] | 0.85 |
| | | 30 | 4.97 | [1.50, 1.80] | 0.47 |
| | | 50 | 5.40 | [1.56, 1.87] | 0.18 |
| SDIPE | 500 | 10 | 0.13 | [1.30,1.71] | 0.95 |
| | | 30 | 0.83 | [1.28,1.75] | 0.95 |
| | | 50 | 1.14 | [1.18,1.79] | 0.96 |
| | 1000 | 10 | 0.01 | [1.36,1.65] | 0.96 |
| | | 30 | 0.27 | [1.33,1.67] | 0.95 |
| | | 50 | 0.33 | [1.29,1.72] | 0.94 |

†Percentage of bias relative to true ATE.

Table 2 presents the simulation results for overall 40% of treatment prevalence. Our proposed method demonstrated consistently low relative bias for 500 sample size, ranging from 0.13% to 1.14% as missingness increased. In contrast, the sensitivity-based MI method yielded substantially higher bias levels, with values escalating from 4.65% to 14.78%. A similar pattern was observed for the larger sample size (n = 1000), where the proposed method maintained bias below 0.5% even at 50% missingness, whereas the sensitivity-based approach showed bias between 4.97% and 5.40%. Confidence intervals for the proposed method remained slightly wide yet stable across conditions, while the intervals from the sensitivity-based method were slightly narrower but accompanied by significantly poorer coverage. Specifically, the coverage probability for the proposed method covered around the nominal 95% level in all settings, ranging from 0.94 to 0.96.

Conversely, the coverage for the sensitivity-based method dropped sharply with increasing missingness: from 0.89 to 0.46 at n = 500, and from 0.85 to as low as 0.18 at n = 1000.

Figure 2 and Figure 3 represent the average bias of ATE of 500 replications under a range of delta values for overall treatment prevalence 20% and 40% respectively for our proposed method. These figures denote the reliability of our estimated ATE through different missingness percentage of partially observed confounder and sample sizes. Though there were some deflections in the graph denoting the change of average bias in estimated ATE, but the magnitudes of those deflections were negligible. Large sample size showed comparatively less deflections denoting comparatively slightly more stability in estimated ATE than the small sample size. This scenario was almost same in the both cases, i.e., for smaller and larger sample sizes. Both the graphs indicated a good reliability and robustness of estimated ATE.

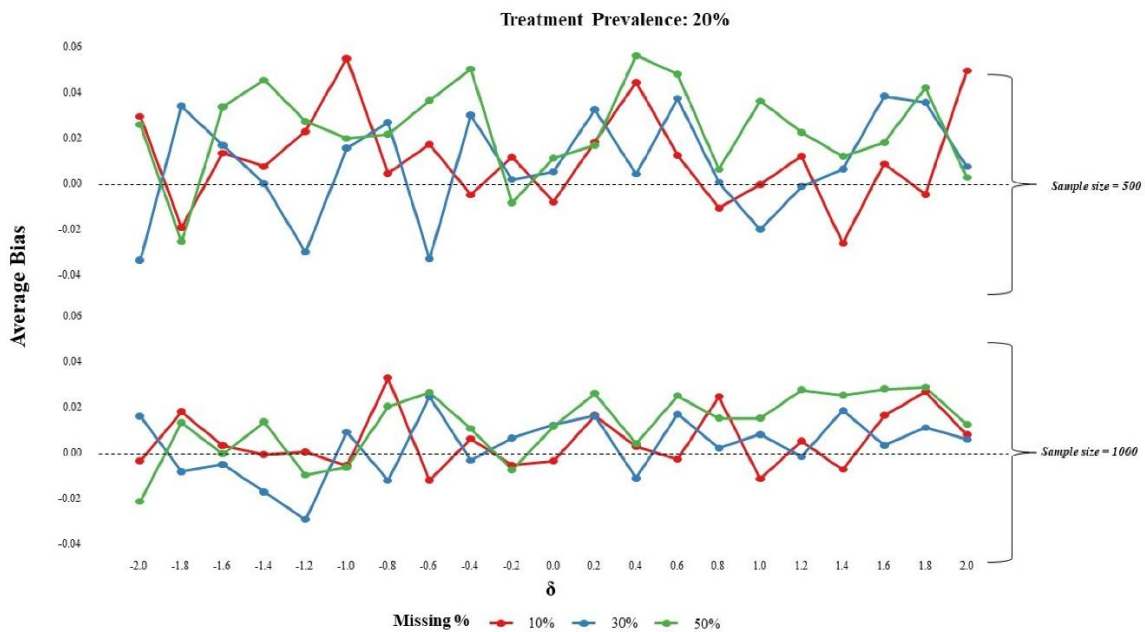

Figure 2 Sensitivity analysis of ATE under treatment prevalence of 20%.

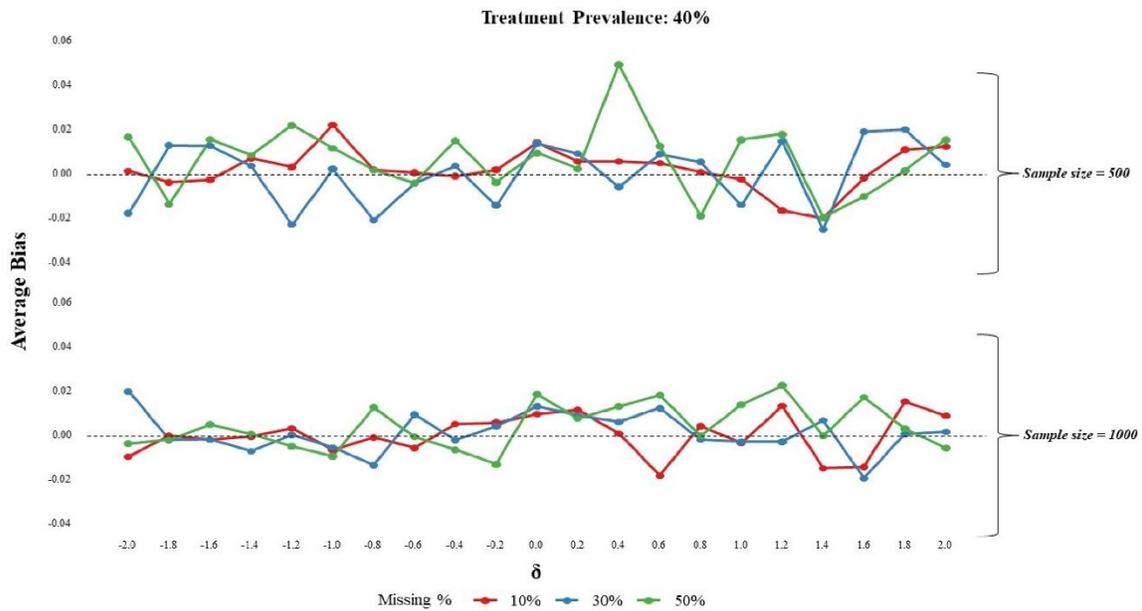

*Figure 3 Sensitivity analysis of ATE under treatment prevalence of 40%.*

**5 Application to Real Data**

We applied our proposed estimation procedure to investigate the causal effect of marital status on depression among adults aged 30 to 55 using data from the 2017-2018 U.S. National Health and Nutrition Examination Survey (NHANES), which is publicly available online. We had a total 1302 individuals of aged between 30 to 55 among which 807 (62%) were married (including living together) and 495 (38%) were single (including divorced or widowed) (Sun & Fu, 2025). Patient Health Questionnaire-9 (PHQ-9) was considered as continuous outcome variable which is a measure of depression severity ranging from 0 to 27 (Kroenke et al., 2001; Sun & Fu, 2025). In 2017-2018 NHANES Questionnaire Data, depressive symptom frequency was assessed by the PHQ-9 instrument. Respondents used this nine-item scale to report depressive symptoms experienced in the past 14 days. We considered Income-to-poverty ratio, age, education level, and gender as potential confounders (Assari, 2020; Kroenke et al., 2001). Only income-to-poverty ratios contained missing values among our confounders. Since individuals with high income may be less likely to disclose their information, it is more likely that missingness in this variable is MNAR (Assari, 2020; Davern et al., 2005; Sun & Fu, 2025). Income data was missing for 10.9% of married individuals versus 11.3% of single participants.

We calculated the ATE along with 95% confidence interval (CI). For CI, 500 bootstrap samples were used. We found that being married was associated with a lower likelihood of depression. Married individuals were found to have a 1.19-point lower PHQ-9 score (95% CI: -1.76, -0.64) on average.

## 6 Discussion

This work addressed the challenge of estimating causal effects in the presence of partially observed MNAR confounders, a setting in which standard identification assumptions fail, and conventional estimators may yield biased results. We proposed a subgroup-specific estimation of causal effects via delta-adjusted multiple imputation and demonstrated consistent performance across varying sample sizes, treatment prevalence, and degrees of missingness relative to a benchmark delta-adjusted multiple imputation approach.

In the recent literature, some work has been done to identify the causal effects in the presence of partially missing MNAR confounders (Lu & Ashmead, 2018; Sun & Fu, 2025; Wen & McGee, 2025). However, one of the major challenges of the developed approaches is the reliance of several strong and unverifiable assumptions, including assumptions like: correct specification of the missingness mechanism, independence of the outcome with the missingness mechanism, independence of the treatment with the missingness mechanism. These assumptions limit applicability of the proposed methods in wider context. Among approaches with fewer assumptions, delta-adjusted MI offers a practical alternative because it does not require the missingness mechanism to be fully known (Dioni et al., 2025). Instead, it assumes that the distribution of missing values systematically differs from the observed values and incorporates this through a user-specified shift parameter ($\delta$) during the imputation process (Jolani, 2012). However, the method's validity depends critically on the correct choice of delta, which must be informed by external data, expert elicitation, or sensitivity analysis (Dioni et al., 2025; Héraud-Bousquet et al., 2012). Additionally, as the distribution of the MNAR confounder differs between observed and missing portion of the data, the relationship between confounders and treatment may also differ across these subgroups. Applying a single propensity score model after imputation assumes a common treatment–covariate relationship, which can leave residual confounding and lead to biased estimates. Our results suggest that estimating the propensity score separately for observed and missing subgroups leads to better covariate balance (Supplementary Table 1 and Supplementary Figure 1 and 2), particularly in the missing subgroup, and therefore can yield more accurate causal effect estimates in the presence of partially observed MNAR confounders. Furthermore, our method demonstrated reduced sensitivity to the choice of the shift parameter (Figures 2 and 3), which highlights the method's robustness to $\delta$ misspecification.

### 6.1 Strength and Contributions

A key strength of this method is its capacity to address MNAR confounders without relying on strong or unverifiable assumptions about the dependencies among treatment, outcome, and missingness mechanisms, unlike existing approaches. A further novelty of the current study is the estimation of the ATE by calculating stabilized inverse probability weights separately for the missing and observed groups, and then combining the estimates based on their proportional representation in the sample. This subgroup estimation of weights ensures the covariate balance between treatment and control group as shown in the results. Additionally, the low sensitivity to the choice of $\delta$, ensures robustness of the approach under misspecification.

## 6.2 Limitations and Future Work

This approach has several limitations. First, although the method shows relative stability across different values $\delta$, choosing an appropriate value remains challenging without external data or expert knowledge. Future work could address this by using a Bayesian framework to incorporate prior information about plausible delta values. Second, the accuracy of the method relies on the correctness of the parametric imputation model; using machine learning models for imputation may improve performance in future applications. The method may also become unstable when the proportion of missing data is extremely high. Currently, the framework is designed for continuous outcomes and a single partially missing confounder. Future extensions could incorporate binary or categorical outcomes, multiple MNAR confounders, and potentially time-varying covariates to broaden the method's applicability.

**Acknowledgments**

Not applicable.

**Ethical Statement**

Not applicable.

**Funding Statement**

The authors received no financial support for the research, authorship, and/or publication of this article

## Data Accessibility

The U.S. National Health and Nutrition Examination Survey (NHANES) data for the 2017–2018 cycle is publicly available through the Centers for Disease Control and Prevention (CDC) (https://wwwn.cdc.gov/nchs/nhanes/default.aspx).

## Competing interests

The authors declared no potential conflicts of interest with respect to the research, authorship, and/or publication of this article.

## Author's Contributions

MNS conceptualized and developed the methodological framework, conducted all simulations and data analysis, and drafted the manuscript. MHP conceptualized the study, conducted all simulations and data analysis, and drafted the manuscript.

## ORCID iD

Md. Niamul Islam Sium: https://orcid.org/0009-0005-1822-8294

Mohammad Hridoy Patwary: https://orcid.org/0009-0008-4920-7017

**Supplementary Table 1.** Comparison of Covariate Balance Between the Proposed Approach and Delta-Adjusted MI (Absolute Average Across 500 Simulations).

| Treatment Prevalence | Missingness | Confounder | Subgroup of Data | Difference in Mean | |
|---|---|---|---|---|---|
| | | | | Proposed approach | Delta adjusted MI |
| 20 | 10 | W | Observed | 0.063 | 0.066 |
| 20 | 10 | W | Missing | 0.148 | 0.311 |
| 20 | 10 | Z | Observed | 0.069 | 0.103 |
| 20 | 10 | Z | Missing | 0.144 | 0.432 |
| 20 | 30 | W | Observed | 0.067 | 0.103 |
| 20 | 30 | W | Missing | 0.090 | 0.148 |
| 20 | 30 | Z | Observed | 0.076 | 0.180 |
| 20 | 30 | Z | Missing | 0.081 | 0.311 |
| 20 | 50 | W | Observed | 0.084 | 0.126 |
| 20 | 50 | W | Missing | 0.063 | 0.091 |
| 20 | 50 | Z | Observed | 0.093 | 0.224 |
| 20 | 50 | Z | Missing | 0.065 | 0.239 |
| | | | | | |
| 40 | 10 | W | Observed | 0.027 | 0.066 |
| 40 | 10 | W | Missing | 0.069 | 0.309 |
| 40 | 10 | Z | Observed | 0.025 | 0.104 |
| 40 | 10 | Z | Missing | 0.074 | 0.435 |
| 40 | 30 | W | Observed | 0.031 | 0.103 |
| 40 | 30 | W | Missing | 0.035 | 0.148 |
| 40 | 30 | Z | Observed | 0.032 | 0.181 |
| 40 | 30 | Z | Missing | 0.037 | 0.311 |
| 40 | 50 | W | Observed | 0.037 | 0.126 |
| 40 | 50 | W | Missing | 0.028 | 0.088 |
| 40 | 50 | Z | Observed | 0.040 | 0.235 |
| 40 | 50 | Z | Missing | 0.027 | 0.247 |

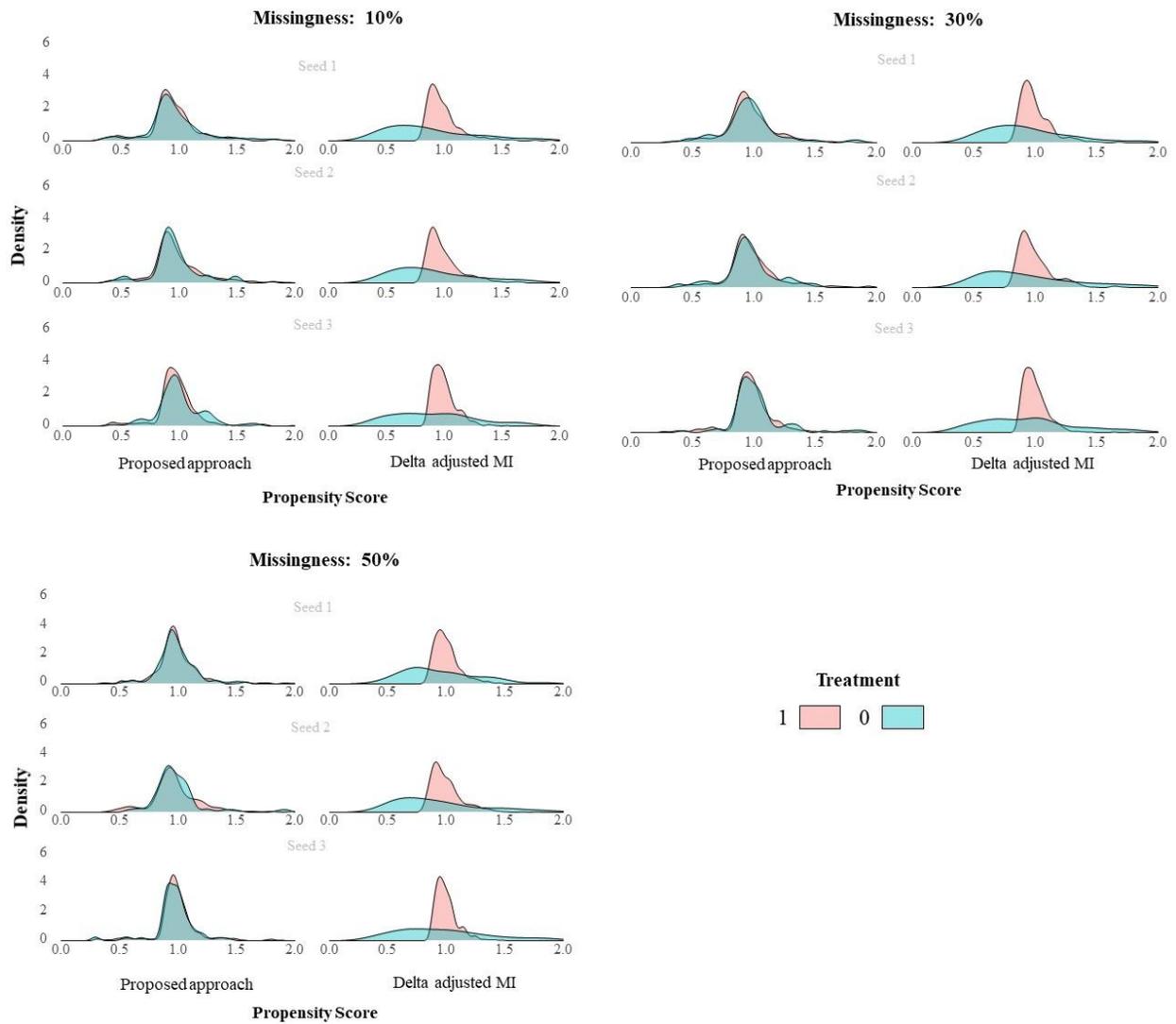

**Supplementary Figure 1.** Comparison of covariate balance between the proposed approach and delta-adjusted MI for treatment prevalence of 20%, missingness levels of 10%, 30%, and 50%, with a sample size of 500, using three random datasets.

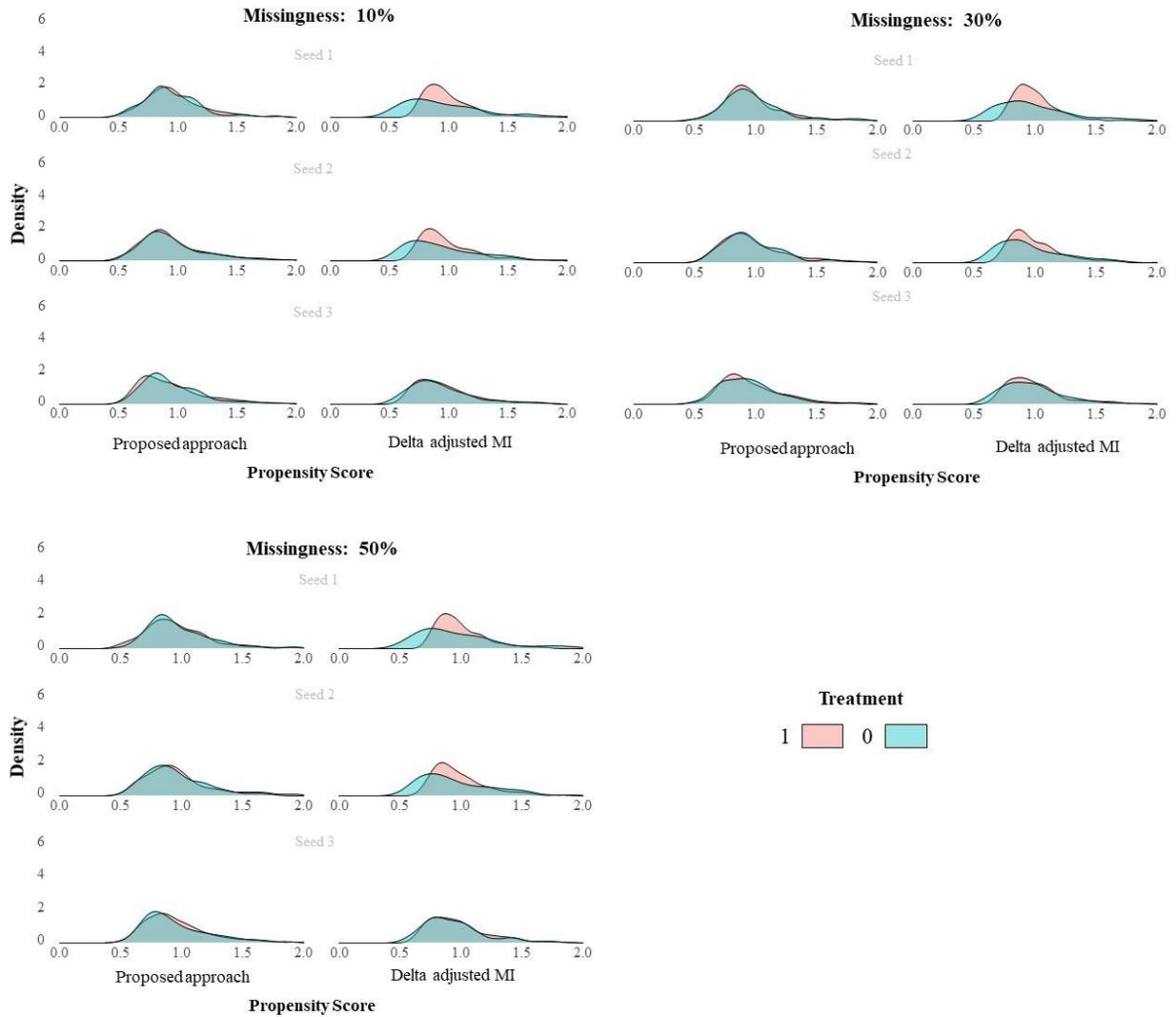

**Supplementary Figure 2.** Comparison of covariate balance between the proposed approach and delta-adjusted MI for treatment prevalence of 40%, missingness levels of 10%, 30%, and 50%, with a sample size of 500, using three random datasets.